# Single Ion Adsorption and Switching in Nano-Electronics

Adam W. Bushmaker*, Vanessa Oklejas, Don Walker, Alan R. Hopkins,
*The Aerospace Corporation*
*Physical Science Laboratory*
*El Segundo, CA*
*USA*

Jihan Chen and Stephen B. Cronin
*The University of Southern California*
*Los Angeles, CA*
*USA*

Single ion detection has, for many years, been the domain of large devices such as the Geiger counter, and studies on interactions of ionized gasses with materials have been limited to large systems[1,2]. To date, there have been no reports on single gaseous ion interaction with microelectronic devices, and single neutral atom detection techniques have shown only small, barely detectable responses[3-6]. Here, we report the first observation of single ion adsorption onto individual carbon nanotubes (CNTs), which, due to the severely restricted one-dimensional current path, experience discrete, quantized resistance increases of over two orders of magnitude. Only positive ions cause changes, by the mechanism of ion potential induced carrier depletion, which is supported by density functional and Landauer transport theory. Our observations reveal a new single-ion/CNT heterostructure with novel electronic properties, and demonstrate that as electronics are ultimately scaled towards the one-dimensional limit, atomic scale effects become increasingly important.

Electronic conduction in one-dimensional channels has been the focus of many research efforts over the past fifteen years, and during that time carbon nanotubes (CNTs) have often served as the prototypical system for studying one-dimensional conduction. The reason for the strong interest is both due to the interesting fundamental physics that occur in such systems[7-9] and because of the enhanced functional performance that has been predicted to occur in CNT-based field effect transistors (FETs) such as high linearity and high operation frequency[10,11]. Also, CNT devices have been developed for new applications using the unique properties of this new material, such as sensors[12], RF NEMs devices[13], and flexible electronics and displays[14]. CNT FETs are now being considered as a promising new technology for next-generation micro- and nano-electronic devices and circuitry.

Due to their small physical size, the conductivity of CNTs and other molecular systems are highly susceptible to changes in charge states of defects nearby or in contact with the material[15], or even by direct chemical activity on the surface of the CNT[16,17]. Similar switching behavior has also been observed in molecular electronic systems[18], and also as random telegraph noise in scaled deep-submicron silicon devices[19]. As dimensions in nano-electronic devices are scaled further and further towards the single-atom, one-dimensional channel limit, effects such as those listed above will become increasingly important, even to the point where they dominate the device operation. In this study, we report the effect of single ion adsorption on the conductance of isolated, suspended, single-walled CNT FETs, which were electrically characterized in-situ during exposure to gaseous ions created by ionizing radiation and high-voltage corona discharge.

Figure 1a shows a scanning electron image of one of the CNT FET devices used in this study. The CNT is suspended over the trench, making electrical contact on either side, with a gate electrode in the bottom of the trench. Figure 1b shows the drain current plotted versus time for the CNT, showing large decreases in drain current observed during exposure to ionized nitrogen gas. The observed transients are characterized by sudden, discrete reductions in current, which had durations ranging from milliseconds to minutes, followed by an equally sudden recovery back to the pristine state. The transient events were observed during exposure to ionized air, Ar, $N_2$, He, and $O_2$, and ceased occurring when the surrounding gas was removed using a vacuum, or when the source of ionization was removed. In Figures 2a and 2b, the device current and resistance during exposure to higher gas ionization rates are plotted, showing multiple, simultaneous switching events, each adding a quantized resistance of ~2.9 G$\Omega$ to the total device resistance. Up to three quantized resistance steps were observed. A cartoon model illustrating multiple ion-defects along the length of the CNT is shown in Figure 2c, and explains this observed behavior. For lower ionization rates and shorter transient lifetimes, multiple steps are not observed, due to the decreased probability for multiple simultaneous interactions. In order to determine the polarity of the ionic species interacting with the CNT, an ionization drift chamber was constructed, and electric fields were applied across the chamber containing the CNT and the surrounding ionized gas. This allowed us to drive either positive or negative ions towards the CNT, depending on the polarity of the electric field. Only positive ions were found to cause the transient switching events to occur, while electrons and negative ions did not.

Data showing the gate voltage dependence of the CNT FET drain current in both the pristine state

(without ions) and transient switched state (during ionized gas exposure) are plotted in Figure 4. Data taken while the CNT was in the pristine state are represented as red circles; data taken while the CNT was exposed to ions are represented by cyan circles, and the dark cyan and dark red lines represent a theoretical transport model, discussed below. The pristine CNT FET shows p-type behavior, with the device being ON for negative gate voltages, with an ON/OFF ratio of ~ $1 \times 10^4$. The subthreshold slope of the drain current in the pristine state is approximately 110 mV/decade. The gate voltage dependence of the drain current from the defected CNT state is similar to that of the pristine state, although the threshold voltage is shifted more negative, and the subthreshold slope is flattened substantially to 400 mV/decade. Both datasets are fit by a theoretical transport model, as outlined below.

The transient switching events observed during exposure to ionized gas are attributed to the ionized gas molecules from the surrounding atmosphere adsorbing onto the CNT and behaving as defects that modify the electrical properties of the carbon nanotube channel. This hypothesis is well supported by the data in Figure 2, which shows discrete quantized resistance increases in the CNT resistance when exposed to ionized gas, and also by the fact that the effect is only observed during exposure to positive ions.

In order to model the effects of adsorbed ionic species on the electrical properties of the CNT, density functional theory modeling was performed using a 1.4 nm long section of an (8,0) semiconducting CNT with and without adsorbed $N_2^+$. First, calculations were performed to find the equilibrium CNT-$N_2^+$ configuration, which gave an equilibrium ion-surface separation of 3.0 Å (ion distance to CNT axis was 6 Å). Results indicate that the binding energy for $N_2^+$ on the surface of the CNT is -11.95 +/- 0.02 eV, which is considerably more favorable than the binding energy for neutral $N_2$, (-0.16 +/- 0.02 eV). Lowdin charge analysis shows that there is also significant charge transfer to the ion of 0.8 electrons from the CNT to the ion. The density of states for the pristine CNT and CNT with adsorbed $N_2^+$ are plotted in Figure 3a, in which the CNT-$N_2^+$ band states are shifted down by ~450 meV. It is hypothesized that despite the presence of free electrons on the CNT and significant charge transfer to the adsorbed ion, the ion does not recombine with an electron to form a free neutral molecule because the total system energy is lower for the adsorbed ion state, as outlined in Figure 3b.

The gate voltage dependence of the switched state was also measured during exposure to ionized gas, and is shown in Figure 4b. The negative threshold voltage shift is caused by the presence of positive charge near the one-dimensional channel, and the decrease in subthreshold slope is characteristic of the lowered effective gate efficiency for the defect. The subthreshold slope for a field effect transistor is described by

$$S = \frac{1}{\alpha}\ln(10)k_B T, \qquad (1)$$

where $\alpha = C_G/C_{Total}$ is the gate efficiency, with gate capacitance $C_G$ and total capacitance $C_{Total}$.

In order to model the effect of the added ion potential on device electrical characteristics with greater fidelity, numerical calculations were performed to solve the discretized quantum mechanical Hamiltonian for the electron eigenstates and quantum mechanical carrier transmission coefficients[20] for the new system consisting of CNT with adsorbed ion potential $U_{Ion}$. Based on these results, a Landauer transport model was used to calculate the gate voltage dependence of the drain current in a CNT with an adsorbed ion, where the potential on the CNT is calculated by solving Poisson's equation[21]:

$$U_{CNT} = \frac{eQ_{CNT}}{C_\Sigma} + e\alpha V_G + U_{Ion}, \qquad (2)$$

where, $Q_{CNT}$ is the charge on the CNT, $V_G$ is the gate voltage, and $U_{Ion}$ is the potential associated with the adsorbed ion, which is given by

$$U_{Ion} = \frac{q_{Ion}}{4\pi\varepsilon_0\sqrt{z^2+h^2}}, \qquad (3)$$

where $h$ is the ion distance from the CNT center axis, $z$ is the distance along the CNT axis relative to the adsorption site, and $q_{Ion}$ is the charge of the ion. The band structure modified by $U_{Ion}$ is shown in Figure 4a, and forms a potential barrier in the CNT valence band, reducing current and shifting the threshold gate voltage towards more negative values. An additional result of this study is that the ion potential forms a single bound electron state near the conduction band edge. The state energy is too high to contribute to device electrical behavior in the p-type ON-state, but the result is nonetheless interesting, and may have important implications for electrical behavior in n-type devices. The nanotube experimental I-$V_G$ data with and without ion exposure was fit with model parameters, and the resulting curves plotted in Figure 4b along with the experimental data. When tunneling effects are taken into account, the estimated value for $q_{Ion}$ based on the fit was +0.18 elementary charges, which agrees with the DFT results showing significant charge transfer to the adsorbed ion. The value for $h$ was found to be 7 Å, which also corresponds well with DFT results for the equilibrium $N_2^+$-CNT axis separation of 6 Å. The defect potential gate efficiency $\alpha_{ion}$ was found to be 14%, which is about 3.5 times lower than the gate efficiency for the device as a whole (50%), and dependent on device electrode geometry. The Landauer transport model with ion potential included fits the data reasonably well, except for in the case of the OFF-state, which is dominated by the noise floor and leakage currents, and above-threshold, where the measured current is lower than that predicted by the model. This discrepancy might be accounted for by defect potential induced charge carrier scattering.

In summary, we report on the extreme sensitivity of nano-electronic devices to single ion adsorption, which is attributed to single-ion induced charge depletion in the one-dimensional channel. Ion adsorption onto the CNT is modeled with density functional theory, which predicts large binding energies, an ion-CNT surface separation of 3 Å, and significant charge transfer. The gate voltage dependence of the CNT current during ion exposure is modeled with a numerical solution of the discretized quantum mechanical Hamiltonian combined with a Landauer electrical transport model, with good agreement between theoretical modeling and experimental results. These experiments have profound impact on our current understanding of the interaction of single ions in one-dimensional nano-electronic materials, and demonstrate the basis for a powerful system to study ion adsorption dynamics[22] at the single-ion level, and also charge transport in CNTs as well as other nano-electronic systems such as single-atom transistors[23], nanowires[24] and quantum dots[25]. Equally important is the demonstration of a new single-ion/CNT based one-dimensional heterostructure, with novel electronic properties such as resonant tunneling and single-electron bound-states. Also, the response of nano-electronic devices to ionized gasses has implications for microelectronic devices operating while exposed to ionizing radiation, as, for instance, in the space radiation environment[26]. The effects observed here raise the prospect of sensitive gas detectors with short exposure times and miniscule detection limits operating at room temperature with almost complete noise immunity. Several important questions remain, such as the cause for the large differences observed in ion-surface adsorption lifetime, and also the nature of interactions between the CNT and ions of different chemical composition.

**Methods** The isolated, suspended CNTs were grown by using chemical vapor deposition (CVD) at temperatures between 800º-900º C, argon bubbled through ethanol as the carbon source, and lithographically defined catalyst islands consisting of a Fe-Mo mixture on an oxide support[27]. After CVD, nearly defect-free, isolated, suspended CNTs are produced, as evidenced by Raman spectroscopy. CNTs span 500 nm deep trench structures with widths from 500 nm – 2 μm, formed in a silicon substrate with 1 μm $SiO_2$/100 nm low stress $SiN_x$ (Figure 1a).

Electrical measurements were performed with an Agilent 4155 semiconductor parameter analyzer, and all measurements were performed at room temperature. Ions where generated using ionizing radiation exposure with 50 MeV protons from the 88" cyclotron facility at Lawrence Berkeley National Lab, and also with gamma radiation (1.17/1.33 MeV) from a Co-60 gamma ray source at The Aerospace Corporation. Ions were also generated using high-voltage corona discharge from commercially available benchtop air ionizers. Density functional theory (DFT) calculations were performed using the plane-wave pseudopotential method implemented by the Quantum ESPRESSO DFT package (v. 5.1)[28], using the spin-polarized PBE gradient corrected functional[29].

Device-level numerical simulations were performed with Matlab™ on an engineering workstation. Eigen-functions of Schrödinger's equation were obtained by solving a discretized Hamiltonian matrix, and energy dependent transmission coefficients through the ion potential barrier were calculated using the propagation matrix method[20]. The details on implementation of the Landauer transport model are reported elsewhere[21,30].

**Acknowledgements** The authors acknowledge Rocky Koga, Jeffrey George, and Steve Bielat of The Aerospace Corporation for assistance with operation of the LBNL 88" cyclotron, and Steve Moss and Robert Nelson of The Aerospace Corporation and Cory Cress of NRL for discussions regarding the interpretation of the data. The portion of the work done at The Aerospace Corporation was funded by the Independent Research and Development Program at The Aerospace Corporation. DFT calculations were performed on a SGI with wall time provided by the DoD HPC Modernization Office. A portion of this work was done in the UCSB nanofabrication facility, part of the NSF funded NNIN network. Sample fabrication was supported by the U.S. Department of Energy, Office of Basic energy Sciences, Division of Materials sciences and Engineering under Award No. DE-FG02-07ER46376.

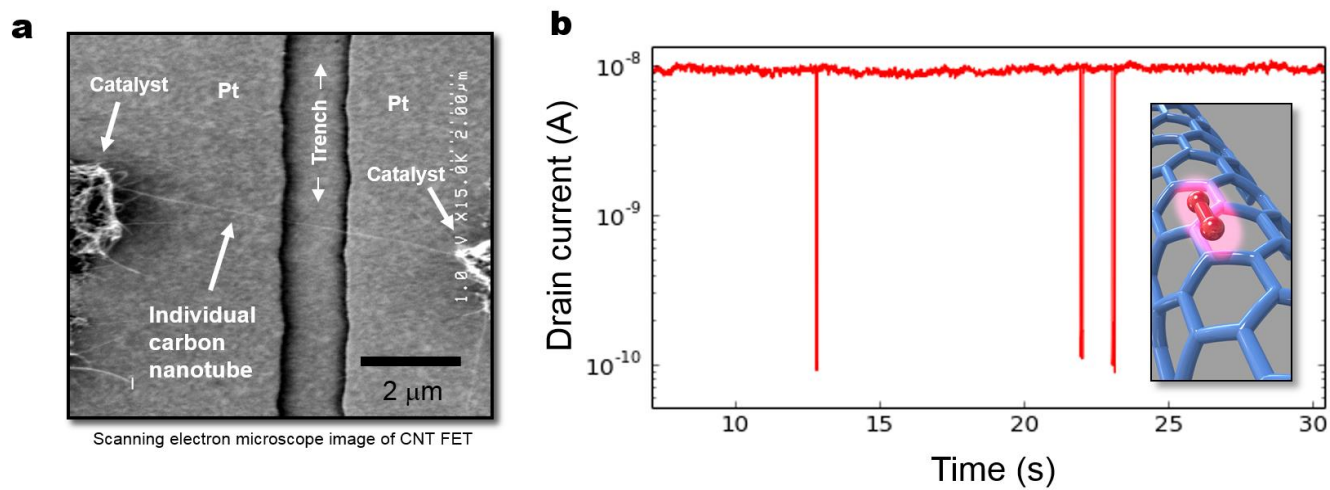

**Figure 1. Device layout and switching transients caused by single ion adsorption. a,** Scanning electron microscope image of CNT FET device and **b,** plot of drain current versus time showing switching transients observed during ionized gas exposure. The inset shows a cartoon image of a gas molecule adsorbed onto the surface of a carbon nanotube.

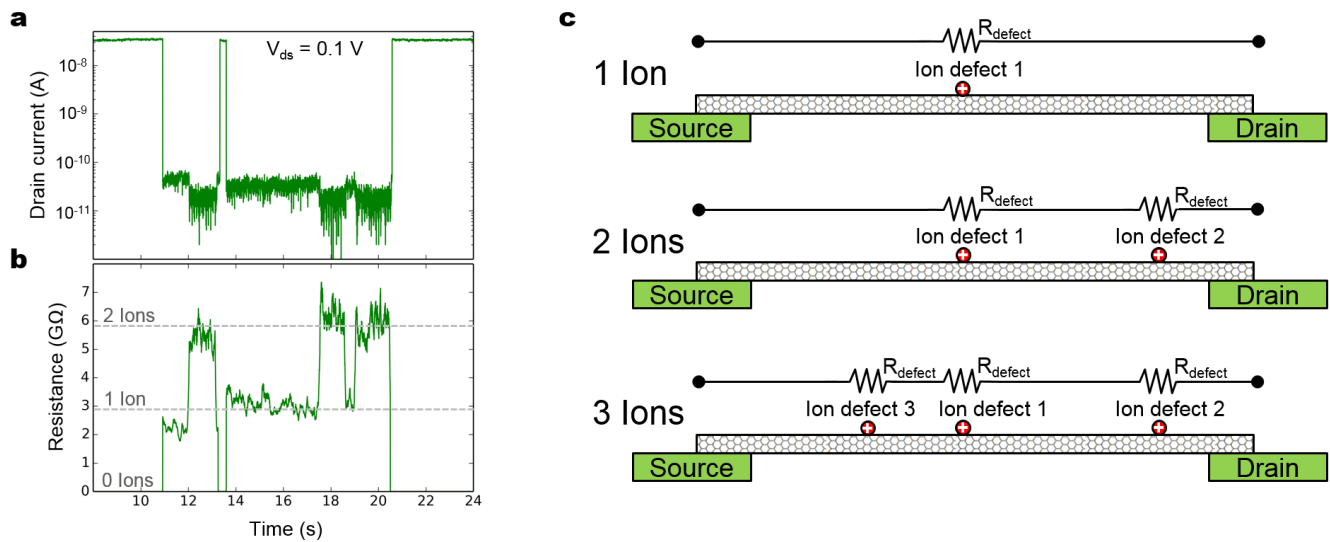

**Figure 2. Multiple simultaneous switching events and cartoon model. a,** Drain current and **b,** resistance plotted versus time illustrating multiple simultaneous switching events occurring during exposure to gaseous ions, and **c,** model for multiple charged defects on surface of the CNT. Up to three simultaneously adsorbed ions were observed.

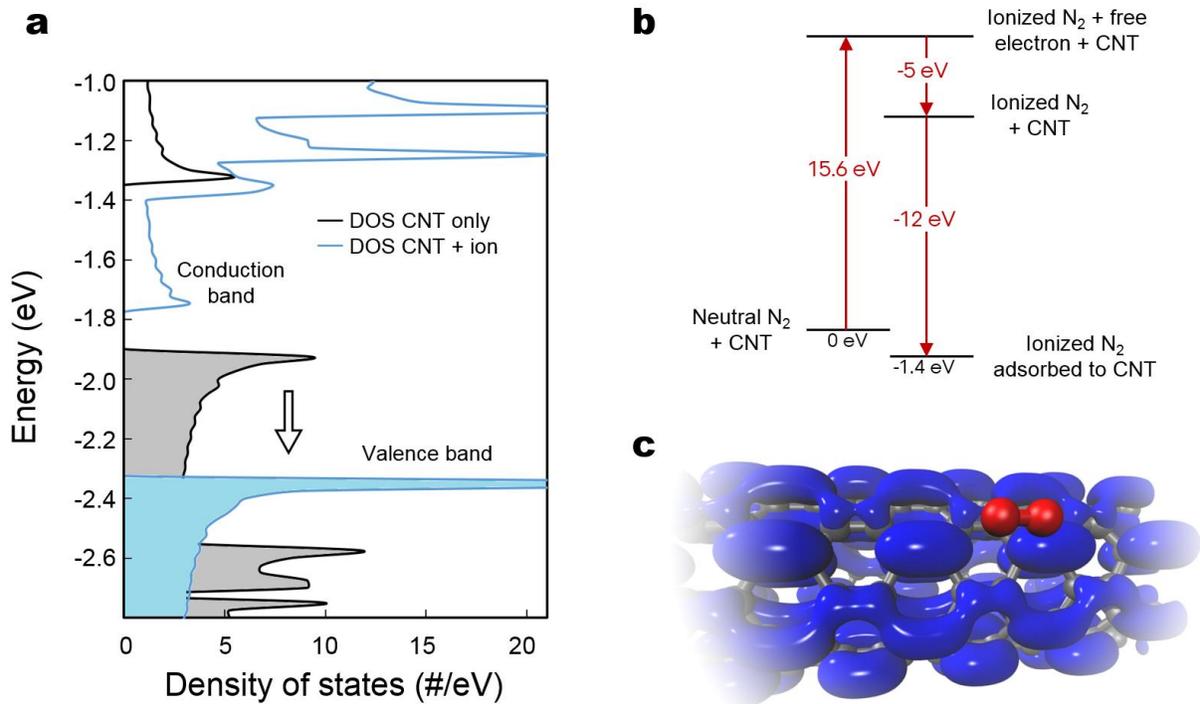

**Figure 3. Density functional theory results. a** Calculated density of states with (cyan) and without (black) adsorbed ion plotted versus energy, **b** ionization/adsorption energy diagram for $N_2^+$ adsorbed to the surface of a (8,0) CNT, and **c** electron density plot for the gamma point conduction band wavefunction on the CNT with adsorbed $N_2^+$ ion. The equilibrium adsorbed ion-CNT separation was found to be ~3 Å, with significant charge transfer of 0.8 electrons to the adsorbed ion. The large binding energies found for ion adsorption onto CNTs indicate quasi-stable ion-residency, which explains the long (milliseconds to minutes) observed lifetimes.

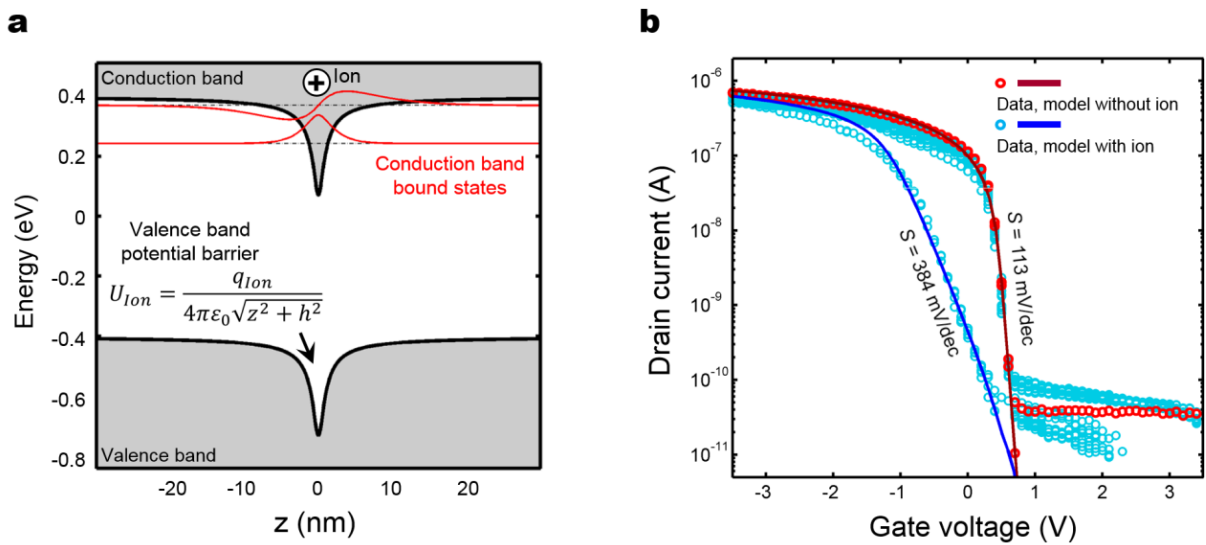

**Figure 4. Ion potential and gate voltage dependence. a** CNT potential plotted versus distance along CNT axis with adsorbed ion, and **b** Drain current plotted versus gate voltage showing experimental data from CNT in both pristine (red circles) and defected (cyan circles) states, along with corresponding results from a Landauer transport model (dark red and dark cyan lines) of the CNT with the adsorbed ion potential shown in **a**. The ion charge $q_{Ion}$ and separation distance $h$ used to calculate the results plotted in **a** and **b** were 0.18 elementary charges and 7 Å, respectively, in concordance with DFT results.